\documentclass[conference]{IEEEtran}

\IEEEoverridecommandlockouts

\usepackage{cite}
\usepackage{amsmath,amssymb,amsfonts}
\usepackage{algorithmic}
\usepackage{graphicx}
\usepackage{textcomp}
\usepackage{xcolor}
\def\BibTeX{{\rm B\kern-.05em{\sc i\kern-.025em b}\kern-.08em
    T\kern-.1667em\lower.7ex\hbox{E}\kern-.125emX}}
    
\usepackage[bookmarks=false]{hyperref}
\usepackage[algoruled,vlined]{algorithm2e}
\usepackage{subfigure}

\makeatletter
\newcommand{\linebreakand}{%
  \end{@IEEEauthorhalign}
  \hfill\mbox{}\par
  \mbox{}\hfill\begin{@IEEEauthorhalign}
}
\makeatother

\begin{document}

\title{Coarse Graining of Data via Inhomogeneous Diffusion Condensation
\thanks{$^*,^\dagger$ Equal contribution. $^{**}$ Corresponding authors. This research was partially funded by: NIH grant F32-NS098616 [\emph{M.W.M.}]; IVADO (l'institut de valorisation des donn\'{e}es) [\emph{G.W.}]; the Alfred P. Sloan Fellowship (grant FG-2016-6607), the DARPA Young Faculty Award (grant D16AP00117), NSF grants 1620216 \& 1912906, NSF CAREER award 1845856 [\emph{M.H.}]; the Chan-Zuckerberg Initiative (grant ID: 182702) [\emph{S.K.}], and NIH grant R01GM135929 [\emph{M.H.,G.W.,S.K.}] The content is solely the responsibility of the authors and does not necessarily represent the official views of the funding agencies.}
}

\author{\IEEEauthorblockN{Nathan Brugnone$^{*}$}
\IEEEauthorblockA{\textit{Dept. of Comp. Math., Sci. \& Eng.} \\
\textit{Michigan State University}\\
East Lansing, MI, USA \\
brugnone@msu.edu}
\and
\IEEEauthorblockN{Alex Gonopolskiy$^{*}$}
\IEEEauthorblockA{\textit{PicnicHealth} \\
Berlin, Germany \\
agonopol@gmail.com}
\and
\IEEEauthorblockN{Mark W. Moyle}
\IEEEauthorblockA{\textit{Dept. of Neuroscience} \\
\textit{Yale University}\\
New Haven, CT, USA \\
mark.moyle@yale.edu}
\and
\IEEEauthorblockN{Manik Kuchroo}
\IEEEauthorblockA{\textit{Interdept. Neurosci. Prog.} \\
\textit{Yale University} \\
New Haven, CT, USA \\
manik.kuchroo@yale.edu}
\linebreakand
\IEEEauthorblockN{David van Dijk}
\IEEEauthorblockA{\textit{Dept. of Internal Medicine} \\
\textit{Dept. of Computer Science} \\
Yale University \\
New Haven, CT, USA \\
david.vandijk@yale.edu}
\and
\IEEEauthorblockN{Kevin R. Moon}
\IEEEauthorblockA{\textit{Dept. of Math. and Stat.} \\
\textit{Utah State University} \\
Logan, UT, USA \\
kevin.moon@usu.edu}
\and
\IEEEauthorblockN{Daniel Colon-Ramos}
\IEEEauthorblockA{\textit{Dept. of Neuroscience} \\
\textit{Yale University} \\
New Haven, CT, USA \\
daniel.colon-ramos@yale.edu}
\linebreakand
\IEEEauthorblockN{Guy Wolf$^{\dagger}$}
\IEEEauthorblockA{\textit{Dept. of Math. and Stat.} \\
\textit{Univ. de Montr\'{e}al; Mila} \\
Montreal, QC, Canada \\
guy.wolf@umontreal.ca}
\and
\IEEEauthorblockN{Matthew J. Hirn$^{\dagger,**}$}
\IEEEauthorblockA{\textit{Dept. of Comp. Math., Sci. \& Eng.} \\
\textit{Dept. of Mathematics} \\
Michigan State University \\
East Lansing, MI, USA \\
mhirn@msu.edu}
\and
\IEEEauthorblockN{Smita Krishnaswamy$^{\dagger,**}$}
\IEEEauthorblockA{\textit{Dept. of Genetics} \\
\textit{Dept. of Computer Science} \\
Yale University \\
New Haven, CT, USA \\
smita.krishnaswamy@yale.edu}
}

\IEEEoverridecommandlockouts
\IEEEpubid{\makebox[\columnwidth]{978-1-7281-0858-2/19/\$31.00~\copyright2019 IEEE \hfill} \hspace{\columnsep}\makebox[\columnwidth]{ }}

\maketitle

\begin{abstract}
Big data often has emergent structure that exists at multiple levels of abstraction, which are useful for characterizing complex interactions and dynamics of the observations. Here, we consider multiple levels of abstraction via a multiresolution geometry of data points at different granularities. To construct this geometry we define a time-inhomogemeous diffusion process that effectively condenses data points together to uncover nested groupings at larger and larger granularities. This inhomogeneous process creates a deep cascade of intrinsic low pass filters on the data affinity graph that are applied in sequence to gradually eliminate local variability while adjusting the learned data geometry to increasingly coarser resolutions. We provide visualizations to exhibit our method as a ``continuously-hierarchical'' clustering with directions of eliminated variation highlighted at each step. The utility of our algorithm is demonstrated via neuronal data condensation, where the constructed multiresolution data geometry uncovers the organization, grouping, and connectivity between neurons. 
\end{abstract}


\section{Introduction}
A fundamental task in data analysis is to characterize variability that separates informative data relations from disruptive ones, e.g., due to noise or collection artifacts. In predictive tasks such as classification, for example, one might seek to extract and preserve information that enhances class separation, while eliminating intra-class variance. However, in descriptive tasks and data exploration, such knowledge does not \textit{a priori} exist, and instead data processing methods must detect emergent patterns that encode meaningful abstractions of the data. Furthermore, it is often the case that data abstraction cannot be conducted at a single scale, and instead one must consider multiresolution data representations that generate several scales of abstraction -- each emphasizing different properties in the data.

The need for multiresolution data representations is of particular importance in biomedical data exploration, where recent technological advances introduce vast amounts of unlabeled data to be explored by limited numbers of domain experts. For example, in single-cell transcriptomics, high-throughput genomic and epigenetic assays have led to an explosion in high-dimensional biological data measured from various systems including imaging~\cite{giesen2014highly,angelo2014multiplexed}, mass cytometry~\cite{Bendall2011}, and scRNA-seq~\cite{shapiro2013single,kolodziejczyk2015technology}. To fully utilize this transformative big data availability, computational methods are needed that leverage the intrinsic data geometry (e.g., using manifold learning techniques \cite{moon2018manifold}) to enable exploratory analysis upon it. 

A common approach towards data abstraction is to use clustering algorithms that provide coarse-grained representations of the data by grouping data points into salient clusters~\cite{levine2015data,galluccio2013clustering,von2007tutorial}, either at a single scale or hierarchically (see Section~\ref{sec: related work}). However, standard clustering algorithms such as $k$-means~\cite{lloyd1982least,kanungo2002efficient} or expectation maximization (EM)~\cite{moon1996expectation} have many limitations. For example, they fail to perform well on high-dimensional data, or they require a number of assumptions about the underlying structure of the data~\cite{Ng2002}. In particular, a primary challenge in clustering is determining the optimal number of clusters or groups. Many algorithms require the user to explicitly choose the number of clusters (as in $k$-means) or tune a parameter that directly relates to the number of detected clusters (e.g., as in Phenograph~\cite{levine2015data}). In exploratory settings, this makes it particularly challenging to detect small, unique, or otherwise rare data type clusters, and extract new knowledge from them.

Here, we present a new approach to address the challenge of multiscale data coarse graining by using a data-driven time-inhomogeneous diffusion process, which we call diffusion condensation. Our proposed diffusion condensation process learns a ``continuous hierarchy'' of coarse-grained representations by iteratively contracting the data-points towards a time-varying data \textcolor{black}{manifold} that represents increasingly coarser resolutions. At each iteration, the data points move to the center of gravity of their local neighbors as defined by this data-driven diffusion process~\cite{Coifman2006,Nadler2005}. This in turn alters the next steps of the diffusion process to reflect the new data positions. Across iterations, this construction creates a time-inhomogeneous Markov process on the data, which represents the changing affinities between data points, along with changing granularities. The process eventually collapses the entire data set to a single point. However, intermediate steps in this process produce coarse-grained data representations at particular granularities or abstraction levels. Importantly, our results show that distinct clusters emerge at different scales and each data point (e.g., each cell in transcriptomic data) is represented by a time series of feature vectors that capture multiresolution information in the data. Therefore, the data embedding provided by the constructed diffusion condensation process can be thought of as a dynamic video, as opposed to static snapshots provided by traditional manifold learning, such as diffusion maps~\cite{Coifman2006} and other dimensionality reduction methods~\cite{van2009dimensionality}.

\section{Related work}
\label{sec: related work}
Typical attempts at providing multiscale data abstraction or summarization rely on hierarchical clustering, which is a family of methods that attempts to derive a tree of clusters based on either recursive agglomeration of datapoints or recursive splitting. Agglommerative methods include the popular linkage clustering, or community detection methods including the Louvain Method~\cite{blondel2008fast}. Splitting based approaches include recursive bisection \cite{dasgupta2006spectral} and divisive analysis clustering \cite{kaufman2009finding}. At each iteration, these methods explicitly attempt to discover the best split or merge at each iteration, thereby forcing points together or apart as the case may be. Diffusion condensation by contrast does not force any splits or mergers at any iteration and simply allows datapoints to come together naturally via repeated condensation steps. Thus, there may be many iterations in which a cluster of datapoints remains distinct from other clusters. This time length under which the cluster persists can itself be a metric of the distinctness of a cluster, and the agglomeration of all such cluster persistence times creates a diagram similar to those created in persistent homology~\cite{wasserman2018topological,kwitt2015statistical}. Thus the hierarchical tree created by diffusion condensation (displayed as a Sankey diagram in Figure~\ref{fig:scRNA-sanky}) has branches whose lengths are meaningful in terms of cluster separation.

\section{Preliminaries}
\label{sec: preliminaries}
\paragraph{Manifold learning}
High dimensional data can often be modeled as originating from a sampling $Z = \{z_i\}_{i=1}^N \subset \mathcal{M}^d$ of a $d$ dimensional manifold $\mathcal{M}^d$ that is mapped to $n \gg d$ dimensional observations $X = \{x_1, \ldots, x_N\} \subset \mathbb{R}^n$ via a nonlinear function $x_i = f(z_i)$. Intuitively, the reason for this phenomenon is that data collection measurements (modeled here via $f$) typically result in high dimensional observations, even when the intrinsic dimensionality, or degrees of freedom, in the data is relatively low. This manifold assumption is at the core of the vast field of manifold learning (e.g.,~\cite{moon2018manifold,Coifman2006,van2009dimensionality,izenman2012introduction}, and references therein), which leverages the intrinsic data geometry, modeled as a manifold, for exploring and understanding patterns, trends, and structure in data.

\paragraph{Diffusion geometry}
In~\cite{Coifman2006}, diffusion maps were proposed as a robust way to capture intrinsic manifold geometry in data using random walks that aggregate local affinity to reveal nonlinear relations in data and allow their embedding in low dimensional coordinates. These local affinities are commonly constructed using a Gaussian kernel
\begin{equation}
\label{GKernel}
\mathbf{K} (x_i, x_j) = \exp\left( {-\frac{\| x_i- x_j\|^2}{\varepsilon}  }\right) \, , \quad i,j=1,...,N
\end{equation}
where $\mathbf{K}$ is an $N \times N$ Gram matrix whose $(i,j)$ entry is denoted by $\mathbf{K}(x_i, x_j)$  to emphasize the dependency on the data $X$. The bandwidth parameter $\varepsilon$ controls neighborhood sizes. A diffusion operator is defined as the row-stochastic matrix $\mathbf{P} = \mathbf{D}^{-1} \mathbf{K}$ where $\mathbf{D}$ is a diagonal matrix with $\mathbf{D} (x_i, x_i) = \sum_j \mathbf{K} (x_i,x_j)$, which is referred to as the degree of $x_i$. The matrix $\mathbf{P}$ defines single-step transition probabilities for a time-homogeneous diffusion process (which is a Markovian random walk) over the data, and is thus referred to as the diffusion operator. Furthermore, as shown in \cite{Coifman2006}, powers of this matrix $\mathbf{P}^t$, for $t > 0$, can be used for multiscale organization of $X$, which can be interpreted geometrically when the manifold assumption is satisfied. 

\paragraph{Diffusion filters} While originally conceived for dimensionality reduction via the eigendecomposition of the diffusion operator, recent works (e.g.,~\cite{van2018recovering,lindenbaum2018geometry, gama:diffScatGraphs2018, gao:graphScat2018}) have extended the diffusion framework of \cite{Coifman2006} to allow processing of data features by directly using the operator $\mathbf{P}$. In this case, $\mathbf{P}$ serves as a smoothing operator, and may be regarded as a generalization of a low-pass filter for either unstructured or graph-structured data. Indeed, consider a vector $\mathbf{v} \in \mathbb{R}^N$ that we think of as a signal $\mathbf{v}(x_i)$ over $X$. Then $\mathbf{P} \mathbf{v} (x_i)$  replaces the value $\mathbf{v}(x_i)$ with a weighted average of the values $\mathbf{v}(x_j)$ for those points $x_j$ such that $\| x_i - x_j \| = O (\sqrt{\varepsilon})$. Applications of this approach include data denoising and imputation~\cite{van2018recovering}, data generation~\cite{lindenbaum2018geometry}, and graph embedding with geometric scattering~\cite{gama:diffScatGraphs2018, gao:graphScat2018}.

\section{Diffusion condensation}
\subsection{Time inhomogeneous heat diffusion}

The matrix $\mathbf{P}$ defines the transition probabilities of a random walk over the data set $X$. Computing powers of $\mathbf{P}$ runs the walk forward, so that $\mathbf{P}^t$ gives the transition probabilities of the $t$-step random walk. Since the same transition probabilities are used for every step of the walk, the resulting diffusion process is time homogeneous.

A time inhomogeneous diffusion process arises from an inhomogeneous random walk in which the transition probabilities change with every step. Its $t$-step transition probabilities are given by
\begin{equation*}
    \mathbf{P}^{(t)} = \mathbf{P}_t \mathbf{P}_{t-1} \cdots \mathbf{P}_1
\end{equation*}
where $\mathbf{P}_k$ is the Markov matrix that encodes the transition probabilities at step $k$. 

Suppose the data set $X$ has an additional parameter $t = 0, 1, 2, \ldots$ that results from measurements $X(t) = \{ x_1(t), \ldots, x_N(t) \}$ of a time-varying manifold $\mathcal{M}^d(\tau)$ at discretely sampled times $\tau_t = \varepsilon t$. Let $\mathbf{P}_t$ be the resulting Markov matrix derived from $X(t)$, constructed according to the anisotropic diffusion process of \cite[Section 3]{Coifman2006} (which is similar to the construction described in Section \ref{sec: preliminaries}). One can show \cite{hirn:timeCoupledDM2014} the resulting inhomogeneous diffusion process $\mathbf{P}^{(t)}$ approximates heat diffusion over the time varying manifold $\mathcal{M}^d(\tau)$. The singular vectors of this process can be used to construct a so-called time coupled diffusion map, which gives time-space geometric summaries of the data $X$.

The perspective of \cite{hirn:timeCoupledDM2014} is that the data is intrinsically time varying. However, one can also start with a static data set $X$ and construct a series of deformations of the data. In this paper we take the latter perspective and deform the data set according to an imposed, data driven time inhomogeneous process $\mathbf{P}^{(t)}$ that reduces variability within the data over time. The resulting process is referred to as condensation, and is described in the next section.

\subsection{The diffusion condensation process}

Recall from Section \ref{sec: preliminaries} that the application of the operator $\mathbf{P}$ to a vector $\mathbf{v}$ averages the values of $\mathbf{v}$ over small neighborhoods in the data. In the case of data $X = \{x_1, \ldots, x_N \} \subset \mathbb{R}^n$ measured from an underlying manifold $\mathcal{M}^d$ with the model $x_i = f(z_i)$ for $z_i \in \mathcal{M}^d$, this averaging operator can be directly applied to the coordinate functions $f = (f_1, \ldots, f_n)$. Let $\mathbf{f}_k \in \mathbb{R}^N$ be the vector corresponding to the coordinate function $f_k$ evaluated on the data samples, i.e., $\mathbf{f}_k (z_i) = f_k (z_i)$. The resulting description of the data is given by $\bar{X} = \{ \bar{x}_1, \ldots, \bar{x}_N \}$ where $\bar{x}_i = ( \mathbf{P} \mathbf{f}_1 (z_i), \ldots, \mathbf{P} \mathbf{f}_n (z_i) )$. The coordinates of $\bar{X}$ are smoothed versions of the coordinates of $X$, which dampens high frequency variations in the coordinate functions and thus removes small perturbations in the data. This smoothing technique is used in \cite{van2018recovering} to impute and denoise data.

Here we consider not only the task of eliminating variability that originates from noise, but also coarse graining the data coordinates to provide multiple resolutions of the captured information in them. Therefore, we aim to gradually eliminate local variability in the data using a time inhomogeneous diffusion process that refines the constructed diffusion geometry to the coarser resolution as time progresses. This condensation process proceeds as follows. Let $X(0) = X$ be the original data set with Markov matrix $\mathbf{P}_0 = \mathbf{P}$ and $X(1) = \bar{X}$ the coordinate-smoothed data described in the previous paragraph. We can iterate this process to further reduce the variability in the data by computing the Markov matrix $\mathbf{P}_1$ using the coordinate representation $X(1)$. A new coordinate representation $X(2)$ is obtained by applying $\mathbf{P}_1$ to the coordinate functions of $X(1)$. In general, one can apply the process for an arbitrary number of steps, which results in the condensation process. Let $X(t)$ be the coordinate representation of the data after $t \geq 0$ steps so that $X(t) = \{ x_1(t), \ldots, x_N(t) \}$ with $x_i(t) = (\mathbf{f}_1^{(t)}(z_i), \ldots, \mathbf{f}_n^{(t)} (z_i) )$, where $\mathbf{f}_k^{(0)} = \mathbf{f}_k$. We obtain $X(t+1)$ by applying $\mathbf{P}_t$, the Markov matrix computed from $X(t)$, to the coordinate vectors $\mathbf{f}_k^{(t)}$. This process results in:
\begin{equation} \label{eqn: condensation}
    \mathbf{f}_k^{(t+1)} = \mathbf{P}_t \mathbf{f}_k^{(t)} = \mathbf{P}_t \mathbf{P}_{t-1} \cdots \mathbf{P}_1 \mathbf{P}_0 \mathbf{f}_k, \quad t \geq 0
\end{equation}

From \eqref{eqn: condensation} we see the coordinate functions of the condensation process at time $t+1$ are derived from the imposed time inhomogeneous diffusion process $\mathbf{P}^{(t)} = \mathbf{P}_t \cdots \mathbf{P}_0$. The low pass operator $\mathbf{P}_t$ applies a localized smoothing operation to the coordinate functions $\mathbf{f}_k^{(t)}$. Over the entire condensation time, however, the original coordinate functions $\mathbf{f}_k$ are smoothed by the cascade of diffusion operators $\mathbf{P}_t \cdots \mathbf{P}_0$. This process adaptively removes the high frequency variations in the original coordinate functions. The effect on the data points $X$ is to draw them towards local barycenters, which are defined by the inhomogeneous diffusion process. Once two or more points collapse into the same barycenter, they are identified as being members of the same cluster. In Section~\ref{sec: properties} we demonstrate condensation's dynamic data deformations to remove variability and collapse points into clusters.
\subsection{Algorithm} 

Pseudocode is provided in Algorithm \ref{alg:condensation}. Although not strictly necessary, cluster convergence may be accelerated by increasing the bandwidth, $\epsilon$, when the $l^{\infty}$-norm of the difference between densities of the previous and current iterations, $\textbf{Diag}(Q')$ and $\textbf{Diag}(Q)$, falls below a threshold.

\begin{algorithm}
	\SetAlgoLined
	\SetKwInOut{Input}{Input}
	\SetKwInOut{Output}{Output}
	\SetKwFunction{Distance}{Distance}
	\SetKwFunction{Merge}{Merge}
	\SetKwFunction{Range}{Range}
	\SetKwFunction{Where}{Where}
	\SetKwFunction{Affinity}{Affinity}
	\SetKwFunction{RowNorm}{RowNormalize}
	\SetKwFunction{RowSum}{RowSum}
	\SetKwFunction{Diag}{Diag}
	\Input{$X \gets $ NxM matrix of N data points, M features\; $\epsilon \gets$ initial filter bandwidth}
	\Output{$X_{t} \gets$ NxM data matrix after t condensations}
	\Begin{
		$i \gets 0$; $i_{prev} \gets -2$\; 
		$Q' \gets I_{N}$\; 
		$Q_{diff} \gets \infty$\;
		$\text{labels} \gets \Range(0,N)$\;
		\While{$ i - i_{prev} > 1$}{
			$i_{prev} \gets i$\;
			\While{$ Q_{diff} >= ~1\times 10^{-4}$}{
				$i \gets i+1$\;
				$D \gets \Distance(X) $\;
				$\Merge(\text{labels}[\Where(D<1\times 10^{-3})])$\;
				$A \gets \Affinity(D)$\;
				$Q \gets \Diag(\RowSum(A))$\;
				$K \gets Q^{-1}A Q^{-1} $\;
				$P \gets \RowNorm(K)$\; 
				$X \gets P \times X $\;
				$Q_{diff} \gets ||\Diag{Q} - \Diag{Q'}||_{l^{\infty}}$\;
				$Q' \gets Q$\;
			}	
			$\epsilon \gets \epsilon \times 2$\;
			$Q_{diff} \gets \infty$\;
		}
	}
	\caption{Condensation \label{alg:condensation}}
\end{algorithm}

The present implementation provides proof-of-concept. We see computational complexity is dominated by matrix multiply and is $\mathcal{O}(n^{4})$ when $t\geq n$. Thus, more research is needed to scale the algorithm.

\section{Properties of diffusion condensation}
\label{sec: properties}
\subsection{Cluster self-organization}
\begin{figure*}[ht]
	\centering
		\includegraphics[width=\textwidth]{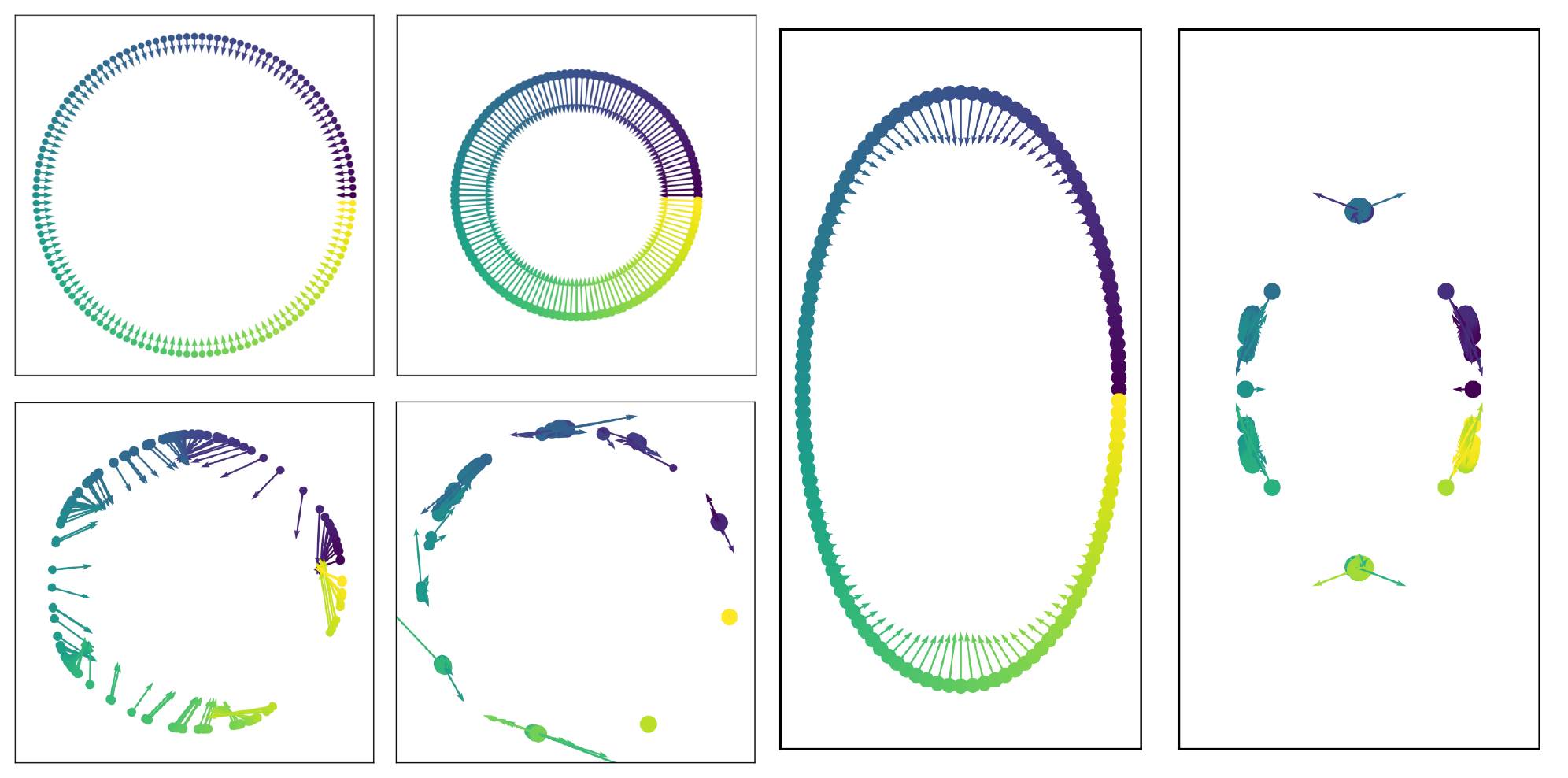}
	\label{fig:f}
	\caption{Condensation of hyperuniform circle (top-left), uniform circle (bottom-left), and hyperuniform ellipse (right) at early/late iterations (left/right, respectively). Point radius corresponds to local density. Arrows computed via the infinitesimal generator $\frac{\mathbf{P}_{k}-I_{N}}{\epsilon}$ show the gradient field and clearly depict data point acceleration during cluster condensation.}
	\label{fig:clusters}
\end{figure*}

Unlike other state-of-the-art clustering algorithms, such as $k$-means, diffusion condensation does not require the user to \textit{a priori} choose a potentially arbitrary number of data clusters to find. Rather, condensation grows self-organizing cluster hierarchies that emerge through local interactions among the data manifold's sampling density and curvature variation.  To disentangle and illustrate such properties, we provide condensation video stills in Figure \ref{fig:clusters}. To begin we highlight the hyperuniformly-sampled (i.e., grid-sampled) circle manifold on the top-left of Figure \ref{fig:clusters}, which demonstrates the base case of homogeneous data density and constant curvature. Note the absence of cluster formation. Comparing this to the hyperuniformly-sampled ellipse on the right of Figure \ref{fig:clusters}, we observe the formation of nontrivial condensation clusters, particularly in the regions of high curvature.

Similarly, the uniformly-sampled circle manifold of constant curvature in the bottom-left of Figure \ref{fig:clusters} exhibits local cluster formation. Hence, we conjecture that nontrivial data density or curvature variation are sufficient conditions for the formation of diffusion condensates.

\subsection{Cluster characterization via spectral decay}
\begin{figure*}[ht]
	\centering
	    \includegraphics[width=\textwidth]{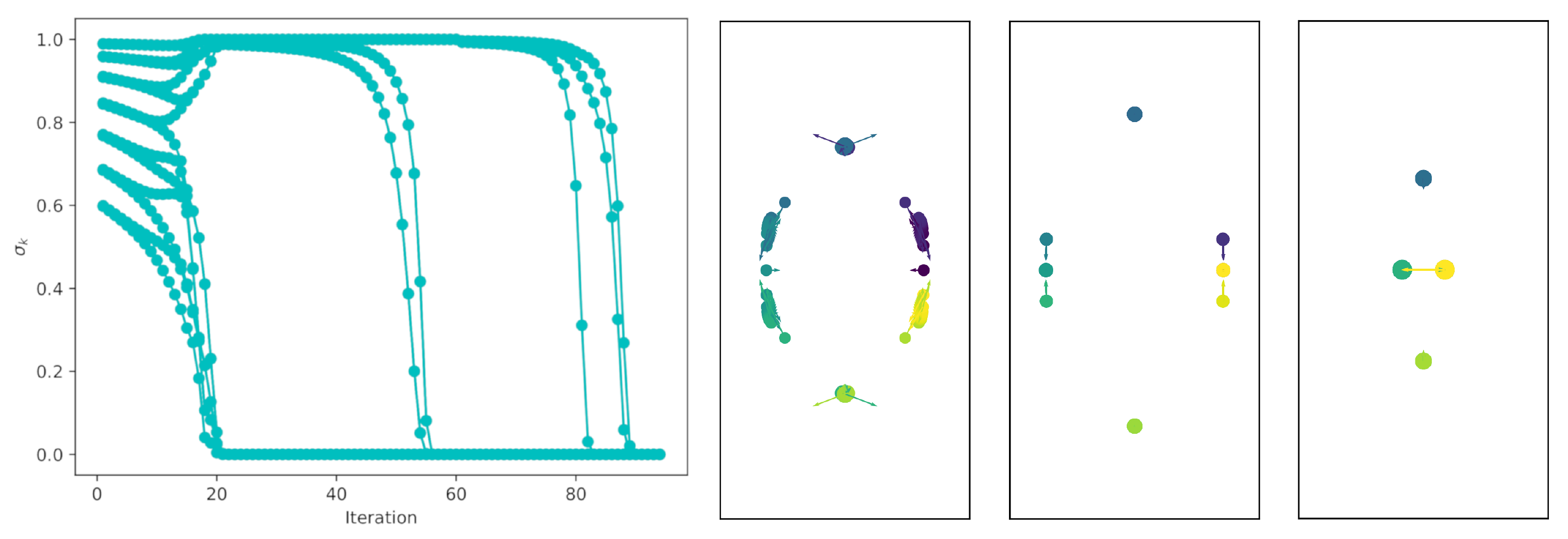}
	\caption{Alternative characterization of hyperuniform ellipse cluster formation via top $14$ nontrivial singular values of the Markov/diffusion operators, $\{\{\sigma_{i}(\mathbf{P}_{k})\}_{k=0}^{t}\}_{i=2}^{15}$ (far left), and corresponding video stills of hyperuniformly-sampled ellipse condensation at iterations $15,53$, and $100$.}
	\label{fig:spectra}
\end{figure*}

\begin{figure*}[ht]
	\centering
	\includegraphics[width=12.5cm]{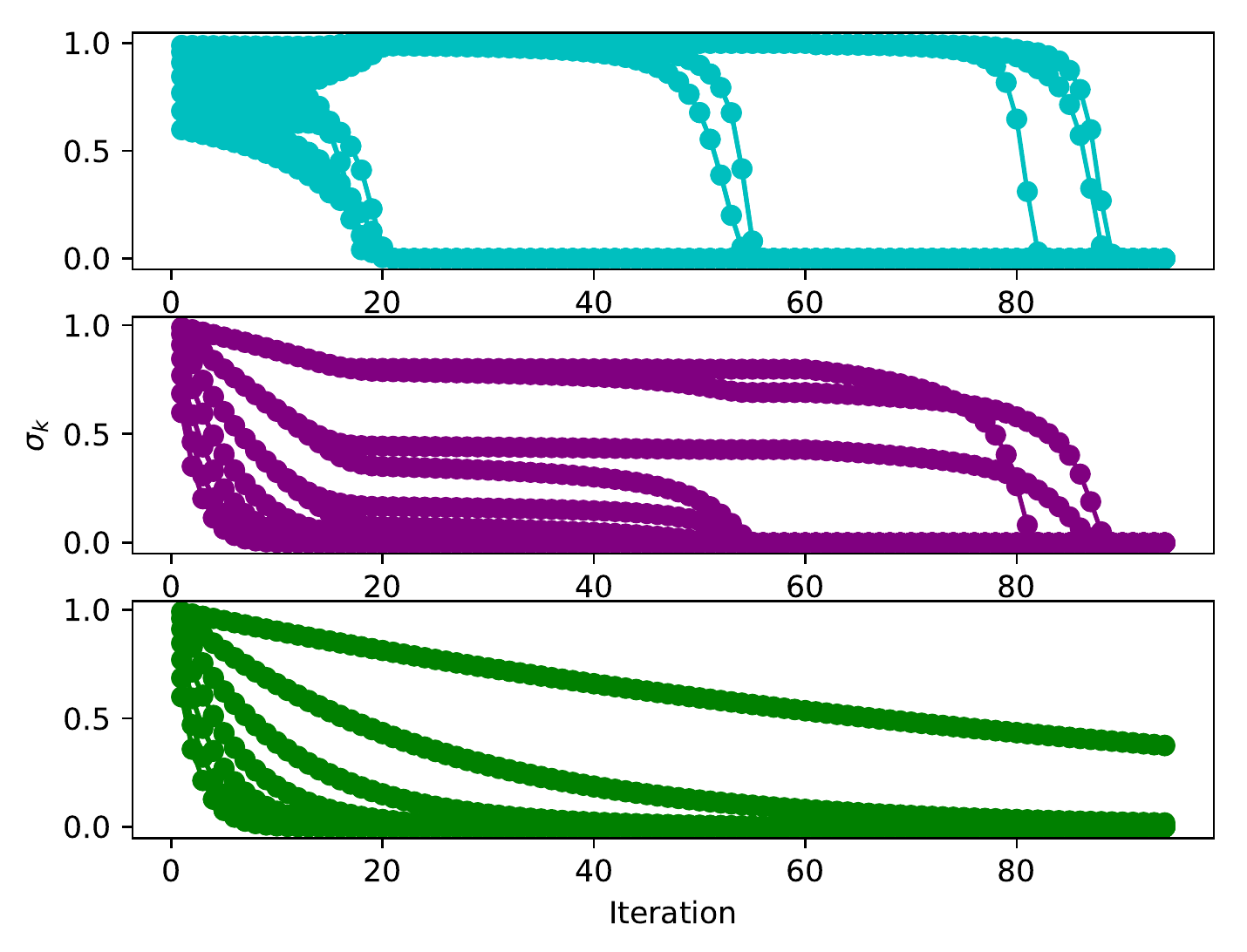}
	\caption{Characterization of hyperuniform ellipse cluster formation via top $14$ nontrivial singular values of the Markov/diffusion operators, $\{\{\sigma_{i}(\mathbf{P}_{k})\}_{k=0}^{t}\}_{i=2}^{15}$ (top, see also Figure \ref{fig:spectra}), $\{\{\sigma_{i}(\mathbf{P}^{(k)})\}_{k=0}^{t}\}_{i=2}^{15}$ (middle), and $\{\{\sigma_{i}(\mathbf{P}^{k})\}_{k=0}^{t}\}_{i=2}^{15}$ (bottom, diffusion maps operator).}
	\label{fig:3-op-spectra}
\end{figure*}

In addition to still frames, it is enlightening to consider cluster formation via its correspondence with the spectral decay. Figure \ref{fig:spectra} demonstrates that data condensation corresponds with sudden, rapid spectral decay. Recall that a nested series of hierarchical data representations may be achieved through diffusion maps by taking successive powers of the diffusion operator, $\mathbf{P}^{t}$ (\emph{not} $\mathbf{P}^{(t)}$), or, equivalently, powering its eigenvalues, $\lambda_{i}^{t}\in[0,1)$ for $i=2,3,\ldots,N$, which function as coordinates of the spectral embedding (e.g., $x_{j} \mapsto \{\lambda_{i}^{t}\psi_{i}(x_{j})\}_{i\geq 2}$, for all $x_{j}\in X$). Of particular interest is the contrast between smooth decay to $0$ of the diffusion maps spectrum as $t\to\infty$ and the rapid, finite-time eigenvalue and singular value decays of $\mathbf{P}_{t}$ and $\mathbf{P}^{(t)}$, respectively, pictured in Figure \ref{fig:3-op-spectra}. The latter characterization may be useful in the identification of hierarchical condensation events in high dimensions, for example. We note that while the condensation operator, $P^{(t)}$, is constructed as in \cite{hirn:timeCoupledDM2014}, its use in clustering is novel. Spectral characterizations of cluster hierarchy persistence further differentiate the present work. For instance, Figure \ref{fig:spectra} displays many features of interest. Most striking is the correspondence between rapid spectral decay of $\mathbf{P}_{t}$ and cluster formation, which are depicted just before the moment of condensation. We see three major areas of cluster formation beginning near iteration 15, again near iteration 53, and once again near the last iteration, 100, when the algorithm halts.

\subsection{Condensation allows multiscale persistence analysis}

Since the condensation algorithm naturally allows points to come together via a low pass filter application at each iteration, the time-point in the process at which clusters naturally come together and the length of time for which a cluster persists (without merging) offer notions of cluster metastability. This can be used to derive a partitioning of the dataspace that has mixed levels of granularity. By contrast, most clustering methods are only able to produce results at a particular granularity; for example, $k$-means tends to favor clusters that roughly divide the data into $k$ partitions of similar sizes. However, different parts of the dataspace may naturally separate at different levels of granularity and this is not visible in other methods. Even hierarchical clustering, due to forced splits and merges, may not reveal the levels of granularity at which data groupings are most distinct. We visualize this persistence information using Sankey diagrams (see Figures~\ref{fig:scRNA-sanky} and~\ref{fig:connectomics}) that show natural groupings of the data. In Section \ref{sec:Rnaseq} we use this capability of condensation to suggest a more relevant subtyping of retinal bipolar neurons on the basis of their transcriptomic profile, as compared to previous literature. 


\section{Empirical results}
\label{sec: empirical}
\subsection{Single-cell transcriptomics data}
\label{sec:Rnaseq}

\begin{figure*}[ht]
	\centering
		\includegraphics[width=\textwidth, height=12cm]{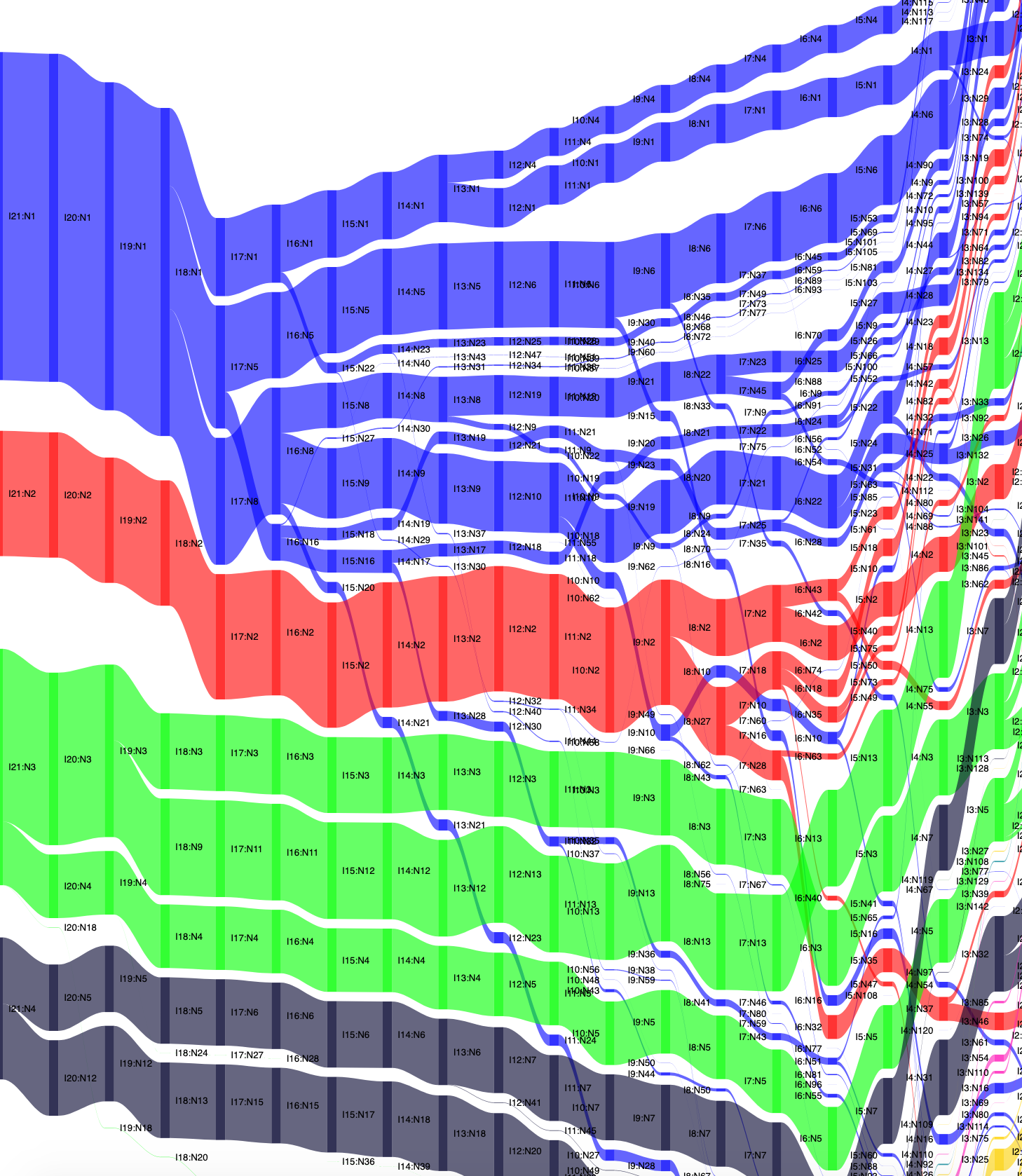}
	\caption{Sankey diagram showing results of 20 iterations of diffusion condensation on the scRNA-Seq retinal bipolar dataset. Left side representing final clusters and right representing earlier stages of the process. The two dark strands represent BC1B and BC1A sub-populations.  Red representing BC3A cell type which becomes distinct quite early in the process. Light green being BC7, forms from three distinct strands suggesting possible subtle sub-populations.
	\label{fig:scRNA-sanky}}
\end{figure*}

A recent study of retinal bipolar neurons using single-cell transcriptomics was performed \cite{Shekhar2016}  to classify cells into coherent subtypes.  The study identified 15 cell sub-types by using the method of \cite{blondel2008fast}, of which 13 were well known and 2 were novel.  We use the findings of said study to benchmark the condensation algorithm.  From the dataset, we use a randomly selected sample of 20,000 cells with gene expression counts sequenced to a median depth of 8,200 mapped reads per cell to perform condensation. The condensation ran for 64 iterations until it achieved a metastable state of 12 clusters (close to the 15 reported in \cite{Shekhar2016}). However due to the continuous clustering history offered by condensation, we are able to assess when these metastable clusters first form; see Figure \ref{fig:scRNA-sanky} for a diagram of iterations $44$ to $64$. A key advantage of the condensation method is its ability to compute cluster persistence based on the lengths of the clustering tree branches, which we use to reassess the subtyping of retinal bipolar cells performed in \cite{Shekhar2016}.

Using community detection methods, \cite{Shekhar2016} found that cluster BC1 (bipolar cone cells, subtype 1) is better described as two clusters, BC1A and BC1B. Shekhar et al. \cite{Shekhar2016} even confirm that morphologically BC1B seems to be a unipolar cluster rather than bipolar.  Condensation clustering corroborates this new finding. Indeed, as shown in Figure \ref{fig:scRNA-sanky}, the dark grey BC1 subclusters stay persistently separated until the last iteration shown. Therefore, the two subcluster-state is more persistent than the single cluster. 

On the other hand, condensation suggests alternative groupings of other clusters not identified by previous papers on retinal bipolar neurons including \cite{Shekhar2016} . Among these, we find that although BC3 has been described in terms of two subcomponents, BC3A and BC3B in biological literature, and in \cite{Shekhar2016}, these subclusters merge early (iteration 53) and the transcriptional profiles are not significantly distinct overall, despite certain selective markers such as Erbb4, Nnat being different between the two. Additionally, we find that our results strongly suggest that BC7 consists of 3 distinct subtypes that persist separately until the last iteration. Previously, the BC7 cell type has been described as a Vstm2b+Casp7+ cone cell that is distinct from other BC types as predicted by \cite{Shekhar2016}. Our analysis, however, reveals that there may be multiple sub-populations that are distinct within this cell type designation. While additional experimentation are required to follow up on this finding, condensation provides a way to examine granularities at which data is best organized based on cluster persistence via the whole condensation history. 
\subsection{Neural connectome data}

\begin{figure*}[!tb]
    \centering
    	\includegraphics[width=\textwidth]{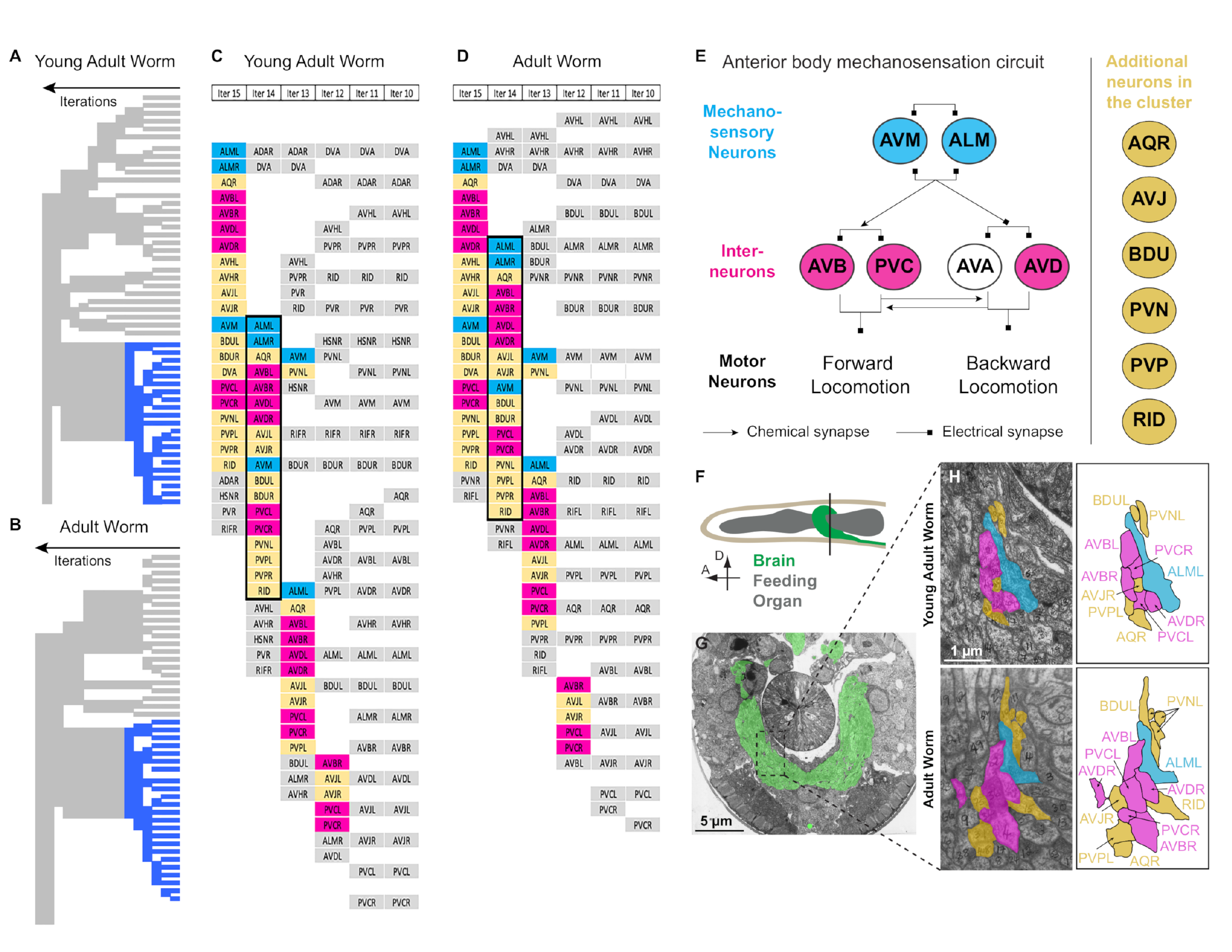}
    \caption{A, B) Sankey diagrams of the condensation results for two \textit{C. elegans} connectomes: a young adult (A) and an adult worm (B). C, D) Sankey diagrams of a subset on neurons (blue in A and B) for a young adult (C) and an adult worm (D). Letter code corresponds to specific neuron names.  Cells are pseudocolored based on the function of the specific neuron in the circuit, with blue representing mechanosensory neurons, red representing interneurons, and yellow representing additional neurons unknown to function in this circuit. Note similar condense profiles at the level of single cells between both young adult (C) and an adult worm (D) connectomes. Cluster outlined in black is the cluster analyzed further in (E).  E)	Circuit diagram for the anterior body mechanosensation circuit. We color the specific neurons according to function as in (C and D).  Circuit diagram adapted from \cite{girard2006wormbook}. F) Cartoon depicting the head of \textit{C. elegans}. The vertical black line shows where the electron micrograph serial section was collected. G) Serial electron micrograph image. Neurons corresponding to the brain of the animal are highlighted green. H) Cropped view of a cross section from the serial electron micrographs corresponding to the anterior body mechanosensation circuit (represented in E). Neurons are pseudocolored as in (C-E). Note how both the relative positions contact profiles of these neurons are similar between both animals, as predicted by the algorithm.}
    \label{fig:connectomics}
\end{figure*}

Since the condensation algorithm operates via a series of diffusion operators, which can be regarded as types of adjacency matrices, we sought to understand if the algorithm would apply to coordinate-free spaces. To achieve this we took a datatype that naturally exists as a graph: the neural connectome data of the \textit{Caenorhabditis elegans} brain, a neuropil called the “nerve ring” consisting of 181 neurons. Here an adjacency matrix was created from the contact profiles determined by images along slices of the worm, i.e., neurons that were more frequently in contact with one another were assumed to have a stronger connection and communication with one another. This adjacency matrix was then eigendecomposed to create a coordinate space in order to perform the condensation. The remainder of the algorithm remained as described.

First we sought to test out the robustness of the condensation algorithm by applying it to two complete connectomes of the \textit{Caenorhabditis elegans} brain.  Previous comparisons between these connectomes had concluded that they largely share similar structure at the level of cell morphology and synaptic positions \cite{white1986structure}. We therefore hypothesized that by comparing the output from these similar connectomes we could test the robustness of our algorithm.  Specifically, we focused on analyzing the relationship between cell-cell contact profiles for every neuron within the two connectomes \cite{brittin2018volumetric}. Cell-cell contact relationships should define modules with the brain that are bundled together, and we hypothesized that if the algorithm was working as expected, it should extract similar contact profiles among the two connectomes. We observed that our algorithm produced similar condense profiles for the two connectomes (See Figure \ref{fig:connectomics}), suggesting that our method can be used to robustly analyze connectomics data. We see that the Sankey diagrams preserve much of the structure including an important mechanosensory circuit. To quantify the similarity between condensation clusters generated from the two connectomes, we compute the adjusted Rand index (ARI) at each condensation iteration from $0$ to $24$ and then take the mean. This yields an $\text{ARI}=0.7$, for $-1\leq\text{ARI}\leq 1$, where the closer to $\text{ARI}=1$ the better.

A major advantage of the diffusion condensation algorithm \ref{alg:condensation} is that it allows analyses of computational iterations to extract biologically relevant information informing the clustering steps. We hypothesized that these iterative steps could reveal units of circuit architecture underlying the brain. To test this, we examined the clusters for well described circuits, specifically, for the anterior body mechanosensation circuit \cite{girard2006wormbook}. The anterior body mechanosensation circuit contains 2 classes of mechanosensory neurons and 4 classes of command interneurons that contact and connect to each other, and based on their contact profile, should be identified by the algorithm \cite{chalfie1985neural, wicks1995integration}.  Indeed, iteration 14 (Figure \ref{fig:connectomics}) identified the circuit in both worms, revealing the predicted relationships between these connecting neurons. Interestingly, iteration 14 also contains neurons of unknown function that, upon closer inspection, are closely associated to the circuit, but have not been implicated in mechanosensory behaviors. Therefore inspection of the condensation algorithm not only extracted the known circuit, but also motivated a new hypothesis regarding the function of unknown neurons associated to the circuit. Together, our analyses demonstrate that the method can be used to compare connectomics data across organisms, to extract biologically relevant units of circuit architecture and even to inform new experiments and discoveries of biological importance. We propose that this method will be broadly useful for systems level analysis of connectomics data. 

\subsection{Algorithm comparison}

\begin{figure*}[ht]
	\centering
		\includegraphics[width=\textwidth]{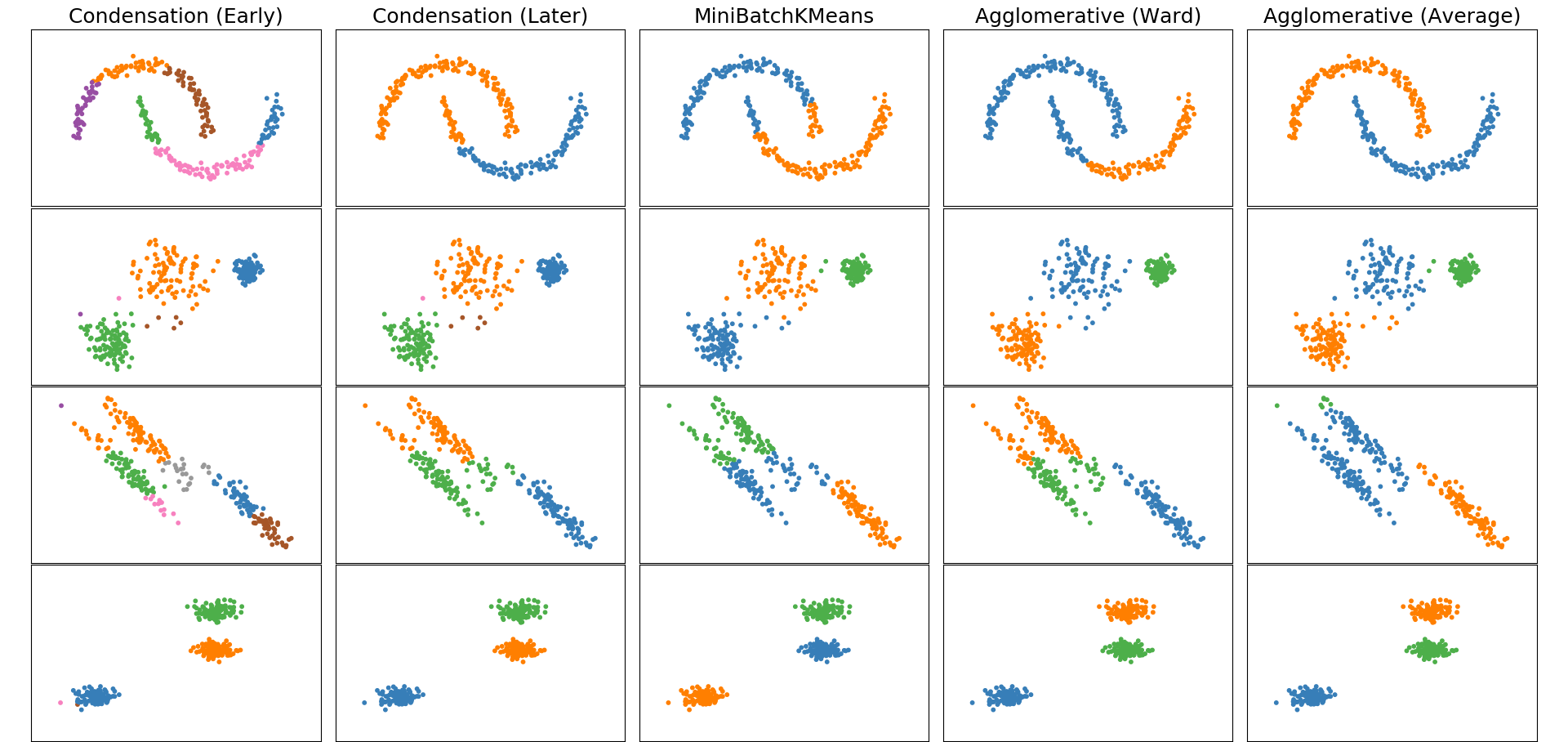}
	\caption{Condensation results as compared with Mini Batch K-Means (center), Agglomerative Clustering with Ward linkage (center-right), and Agglomerative Clustering with average linkage (right) on the scikit-learn clustering dataset ($N=300$). The early condensation snapshot (left) is taken at half the iterations of the later (center-left).}
	\label{fig:compare}
\end{figure*}

We compare condensation at two times with Mini Batch K-Means, Agglomerative Clustering with Ward linkage, and Agglomerative Clustering with average linkage. The two condensation times are early and later, where early is half the iterations of later. The computational experiments are conducted on part of the scikit-learn clustering dataset with default, tuned parameter values and datasets. The variance of the center blob in the Gaussian blobs dataset (Figure \ref{fig:compare}, row two) was decreased from $2.5$ to $1.5$ for separability.

In Figure \ref{fig:compare} we see the earlier iteration of condensation exhibits finer clustering by curvature than the later. Similarly, row three of Figure \ref{fig:compare} exhibits coarser clustering in the later condensation labeling. These examples demonstrate the multiscale nature of clusters assigned via condensation. We note that while we employ only the Euclidean metric in these examples, preliminary tests using other metrics yield promising results. 

\section{Conclusion}
We presented a multiresolution data abstraction approach based on a time-inhomogeneous diffusion condensation process that gradually coarse grains data features along the intrinsic data geometry. We demonstrated the application of this method to biomedical data analysis, in particular in single cell transcriptomics. Furthermore, the presented diffusion condensation can be seen as a cascade of data-driven lowpass filters that gradually eliminates variations in the data to extract increasingly abstract features. Indeed, under this interpretation, the abstraction provided by the condensation process can be related to common intuitions of features extracted by hidden layers of deep convolutional networks, e.g., in image processing. Such features are commonly considered as increasing in abstraction capabilities together with the depth of the network. However, we note that while convolutional networks typically employ relatively-simple pointwise nonlinearities, here the nonlinearity we employ is the reconstruction of the diffusion geometry based on the coarse grained features along the cascade. Therefore, our cascade both learns a multiresolution data geometry and extracts multiresolution characterizations of groupings based on invariant features at each iteration. Finally, we note the increasing interest in geometric deep learning, which aims to tie together filter training in deep networks with non-Euclidean geometric structures that often exist intrinsically in modern data. While our approach here relies only on lowpass filters, it opens interesting directions in employing trained filters (or even designed diffusion wavelets, as done in diffusion and geometric scattering~\cite{gama:diffScatGraphs2018,gao:graphScat2018}) together with geometric reconstruction for multiscale feature extraction from data.

\subsubsection*{Software and media}
All code along with selected media are available at \url{https://github.com/matthew-hirn/condensation.git}.

\bibliographystyle{IEEEtran}
\bibliography{main_camera_ready}

\end{document}